\documentclass[referee]{aa}  
\usepackage{graphicx}
\usepackage{txfonts}

\usepackage{natbib}
\input{psfig}

\begin{document}
%

    \title{Comparing different approaches to model the  rotational modulation of the Sun as a star} 

    \titlerunning{Modelling the rotational modulation of the Sun}


   \author{A.~F.~Lanza
          \inst{1}
          \and
          A.~S.~Bonomo\inst{1,2}
          \and
          M.~Rodon\`o\inst{1,2}\fnmsep\thanks{Deceased}
          }

   \offprints{A. F. Lanza}

   \institute{INAF-Osservatorio Astrofisico di Catania\\
              \email{nlanza@oact.inaf.it}
         \and
             Dipartimento di Fisica e Astronomia, Universit\`a
             degli Studi di Catania, Via S. Sofia, 78, 95123 Catania, Italy\\
             \email{aldo.bonomo@oact.inaf.it}            
             }

   \date{Received ... ; accepted ... }

\abstract{The space missions MOST, COROT and Kepler are going to provide us with high-precision optical photometry 
   of solar-like stars with time series extending  from tens of days to several years. They can be
   modelled to obtain information on stellar magnetic activity by fitting the rotational modulation
   of the stellar flux produced by the brightness inhomogeneities associated with photospheric active regions.    }
   {The variation of the total solar irradiance provides a good proxy for those photometric time series and can be used to 
   test the performance of different spot modelling approaches.}
   {We test discrete spot models as well as maximum entropy and Tikhonov regularized spot models by comparing the reconstructed
   total sunspot area variation and longitudinal distributions of sunspot groups with those actually observed in the Sun
   along activity cycle 23. Appropriate statistical methods are introduced to measure 
   model performance versus the timescale of  variation.}
   {The maximum entropy regularized spot models show the best agreement with solar observations reproducing the total sunspot
   area variation on time scales ranging from a few months to the activity cycle, although the model amplitudes are affected 
   by systematic errors during the minimum and the maximum activity phases. The longitudinal distributions
   derived from the models compare well with the observed sunspot group distributions except 
   during the minimum of activity, when
   faculae dominate the rotational modulation.  The resolution in longitude attainable through
   the spot modelling  is  $\sim 60^{\circ}$, on average. }
   {The application of the maximum entropy modelling to solar analogues will provide us with a quite detailed picture of
    their photospheric magnetic activity that can be the base for comparative and evolutionary studies 
    of solar-like magnetic activity and irradiance variations. }


   \maketitle
%

\section{Introduction}

The Total Solar Irradiance (TSI) has been monitored for almost three decades
by  radiometers that obtained 
a measure of the bolometric disk-integrated flux of our star.
 It shows a complex variability with timescales
ranging from a few minutes to the eleven year cycle, the origin of which is
still not completely understood \citep[see, e.g.,][]{FrolichLean04}.
The peak-to-peak relative amplitude of the long-term change along the activity cycle is of
about 1000 parts per million (ppm) with the maximum irradiance
attained at the maximum of activity, whereas the transits of 
the largest sunspot groups across the solar disk produce irradiance
dips with amplitudes up to 3000 ppm  that last for $10-15$ days.
 
Recent investigations have shown that at least 90\% of the 
irradiance variability on time scales from a few days up to the
11-yr cycle can be accounted for by the
brightness inhomogeneities produced by surface magnetic fields, 
specifically by cool sunspots and warm faculae
\citep[e.g., ][]{Krivovaetal03,deTomaetal04,Wenzleretal05}. 

The study of the solar irradiance variability is limited by the
lack of similar observations for other main-sequence late-type stars
because the current photometric accuracy from the ground is only of 
$\sim 0.01$ mag (i.e., $\sim 10^{4}$ ppm), allowing us to study the variability only for 
objects younger than a few hundreds Myr \citep[e.g.,][]{MessinaGuinan02}. Only for the seasonal
averages of the optical magnitudes, it is possible to reach an
accuracy of the order of a few $10^{3}$ ppm from the ground 
\citep[cf., e.g., ][]{Radicketal98}. 

The situation is going to change thanks to the space-borne photometric 
experiments already in operation (MOST) or planned for the near
future (COROT and Kepler). The Microvariability and Oscillations of STars (MOST)
satellite can reach a photometric 
accuracy of $50-100$ ppm in 100-min integration time
on bright stars ($3<V<6$) observing a single target 
in a passband between 350 and 700 nm 
for time intervals up to 30-40 days 
\citep{Walkeretal03,Crolletal06}. 

The space mission COROT (COnvection,
ROtation and Transits) will simultaneously observe 
between 6,000 and 12,000 
stars with $12<V<16$, in selected fields of view, 
for time intervals up to 150 days 
to look for planetary transits. 
The time series obtained will have a sampling of 8 min and
will reach a photometric accuracy of $100-200$ ppm in 
one hour integration time for stars of apparent magnitude $V \sim 12-13$.
For stars brighter than $V=14$, COROT will also recover information
on the spectral distribution of the optical flux. For fainter stars, an integrated
measure in the passband $350-1000$ nm will be available \citep{Baglin03,Moutouetal05}.

Kepler is a mission that will continuously observe a field in the constellation of
the Cygnus for at least 4 years, simultaneously monitoring the
flux variations in the passband $400-900$ nm for
about 100,000 stars with a 15-min time sampling. The
photometric accuracy is of 20 ppm for an integration time of 
6.5 hours on a G2V star of $V=12$ \citep{Boruckietal04}. 

Although the photometry obtained by such experiments
will be limited to the optical band, they will provide us with data to 
investigate the irradiance variations of a large
sample of solar-like stars. Photospheric cool spots and warm faculae are expected
to dominate variations in the optical passband, as it is indeed 
observed in the Sun, although a significant contribution to the total
irradiance variation of our star comes also from the ultraviolet 
domain \citep{Unruhetal99,Krivovaetal03}. 

In the case of the Sun, it is possible to model the irradiance variations
by summing up the flux perturbations produced by each active region 
observed on the solar disk at a given time 
\citep[see, e.g., ][]{Chapman87,Fliggeetal98,Krivovaetal03}.
The same approach cannot be applied to distant stars because of the lack of spatial
resolution. Tomographic
techniques based on high-resolution spectroscopy, such as the classic Doppler imaging method,
are applicable only to
fast rotators \citep[${\rm v} \sin i \geq 20-25$ km s$^{-1}$; cf., e.g.,][]{Strassmeier02}.

In the case of slowly rotating solar analogues, the modelling of the irradiance variations
must be based on the information on the location and area of the surface active regions
provided by the rotational modulation of the stellar flux.  
Assuming that the active regions are stable during their transit across the
disk of a star as it rotates,
the variability of the optical flux can be attributed
to the modulation of the visibility of  the active region themselves. 
In the framework of this hypothesis, we can obtain a map of the filling factor of the 
active regions over the surface of a star, independently of its rotation period, 
as it has been done for several very active late-type stars.
Earlier models were based on two circular spots
whose co-ordinates and radii were adjusted in order to fit the light curve rotational
modulation \citep[see, e.g.,][]{Rodonoetal86}. The same kind of approach,
but using three active regions containing cool spots and bright faculae in
a fixed proportion, was applied by \citet{Lanzaetal03} and \citet{Lanzaetal04}
to fit the total as well as the spectral solar irradiance variations. 
The position of the three model active regions and the variation of their
total area along a significant fraction of activity cycle 23 turned out to
be in general agreement with those of the largest observed active regions and
complexes of activity dominating the solar irradiance variations. However, in
a significant number of cases, the longitudinal distribution of the observed
active regions was too much complex to be described with
only three active regions and this resulted in a poor agreement between the
model and the observations.

A similar situation arises also in the case of the most active stars. 
The surface maps obtained from the light curve modelling of the K1 subgiant 
component of the RS CVn system HR 1099 can be compared with the much more detailed
Doppler imaging maps revealing that the spot models based on
two or a few spots are generally not suited to reconstruct the complex 
distribution of surface brightness inhomogeneities characterizing a real active star
\citep{Vogtetal99}. On the other hand, the application of the Maximum Entropy (hereinafter ME)
and Tikhonov (hereinafter T) regularization techniques to a continuous distribution of
spots, showed a significantly better
agreement between the maps based on wide-band photometric data and those
obtained by Doppler imaging. In particular, the longitude distribution of
the spotted area and its total variation versus time can be safely reconstructed
by means of those more sophisticated light curve inversion techniques, as it
was shown by \citet{Lanzaetal06}.  

In the present paper, we apply the ME and  T regularizations 
to model the total solar irradiance variations treating the Sun as a distant 
active star, i.e., without introducing any information coming from its spatially resolved
observations. The resulting maps of the active region distributions will be
compared with the location and the area of the observed sunspot groups to assess
the advantages and drawbacks of the proposed spot modelling techniques. A comparison
with the results obtained with the previous modelling approach of \citet{Lanzaetal03}
will also be presented. 

In this way, we shall test the performance of light curve inversion techniques
to be applied for the analysis of the forthcoming high-precision 
photometric observations of main-sequence late-type stars.  Mapping the distribution of 
their surface brightness inhomogeneities, although with a limited spatial 
resolution, will be of fundamental importance to understand the basic 
mechanisms of solar-like activity and variability along the main sequence.
This will provide us with a new perspective to consider the fundamental problems of
the long and short-term solar irradiance variations, with their impact on the Earth environment
and climate, and the physical processes responsible for magnetic field
amplification and modulation in the Sun and solar-like stars, i.e., the stellar 
hydromagnetic dynamo.

\section{The time series of the total solar irradiance}

We shall model the time series of the TSI obtained by the VIRGO experiment on
board of the satellite SoHO.
VIRGO  has two radiometers to measure the
total solar irradiance, DIARAD and PMO6V,
each of which has a replica that is exposed to the solar radiation only
occasionally to determine the degradation of the one that is in continuous
operation. Details on the measurement principle and performance can be found in
\citet{Frolichetal95} and \citet{Frolichetal97}. The measurements
are first corrected for all the known instrumental and environmental
effects, and reported to 1 AU distance and zero radial velocity, giving the
so-called level 1.0 data. The corrections for the instrumental degradation
due to the exposure to solar radiation and for long-term non-exposure dependent
effects are obtained by comparing the time series of DIARAD and PMO6V 
with each other and with other experiments simultaneously measuring the TSI.
The procedure followed to derive such corrections is described in 
the VIRGO web site\footnote{http://www.pmodwrc.ch/pmod.php?topic=tsi
/virgo/proj\_space\_virgo\#VIRGO\_Radiometry}
and detailed in \citet{Anklinetal99}, \citet{FrolichFinsterle01}, \citet{Frolich03}
and \citet{Crommetal04}. In this way, the final TSI time series of level 2.0 is obtained.
We retrieved it from the file: virgo\_tsi\_h\_v6\_90.dat that is accessible
from the VIRGO web site at the Physikalisch-Meteorologisches Observatorium
Davos -- World Radiation Center. It consists of hourly mean values that range from
7 February 1996 to 28 September 2005 with occasional gaps the length of which is
of the order of a few hours or days, except for the four-month gap that  occurred in
mid 1998, when the spacecraft tracking was lost. The standard deviation of the
hourly TSI data is 20 ppm. 

\section{Models of the solar irradiance variation}
\label{model}

The reconstruction of the surface brightness distribution from the
inversion of disk-integrated flux values, modulated by stellar rotation,
is an ill-posed problem because of the non-uniqueness of the solution
and its instability, i.e., small fluctuations in the input data can produce
large changes in the  map corresponding to the minimum of
the $\chi^{2}$. This situation arises because the
rotational modulation of the flux is a function of the
variation of the surface brightness  versus the longitude,
while it contains very little information on its latitudinal dependence.
When the inclination of the rotation axis is close to $90^{\circ}$, as in the
case of the Sun, the duration of the transit of a 
surface feature across the disk is independent of the latitude (the
effect of the surface differential rotation is negligible in this context)
and, therefore, no information on its latitude can be extracted from the 
light curve. 

The only quantities that can be safely derived by modelling the rotational
modulation of the TSI are the longitudinal distribution of the
covering factor of the active regions and the variation of their total area,
measured with respect to a reference value corresponding to a given value of the
irradiance that we assume to be that of the unperturbed Sun
\citep[cf., e.g.,][]{Lanzaetal03}. The contrast of the surface brightness
inhomogeneities must be assumed because their temperatures cannot be derived from measurements
in a single spectral band. Only when the simultaneous flux modulations in different passbands 
can be compared, we can estimate the temperatures of the spots and  faculae and derive their contrasts 
\citep[cf., e.g.,][]{Lanzaetal04,Messinaetal06}.

In order to overcome the non-uniqueness and the instability of the 
spot modelling based on light curve inversion, two approaches have been
proposed. The first adopts a simple configuration of the brightness 
inhomogeneities, e.g., it assumes that there are only three active regions, and 
derives their co-ordinates and areas by fitting the light curve. In the case of the
Sun as a star, the rotation period is also included among the free parameters
because the short lifetime of the active regions and their random appearance on the
solar disk, especially during the phases of high activity of the 11-yr cycle,
prevent the precise determination of the rotation frequency from the power spectrum of the 
TSI variation  \citep[see][]{Lanzaetal03,Lanzaetal04}. 

This modelling approach has the advantage that a limited number of free
parameters appears in the model, so that the uniqueness and instability problems
are significantly reduced and a meaningful solution can be obtained by minimizing the
$\chi^{2}$. 

A second approach makes use of regularizing functionals, i.e., of
some a priori assumptions on the statistical properties of the brightness map,
to derive a unique and stable map of the surface features. In this case, the
surface of the star is subdivided into a large number of elements and the
covering factors of the active regions within the elements are considered as  
free parameters. Among the potentially infinite solutions that fit the rotational
modulation of the flux, those are selected that maximize the configurational
entropy  (ME solution) or minimize the Tikhonov functional (T solution)
of the map. 
In this case, the rotation period needs to be fixed at the outset,
because of the large number of free parameters that would make its
variation impratical in terms of computing time. 

The implementation of these approaches is described in some detail below.

\subsection{The three-spot model of the TSI variation}
\label{3spots}

The model is described  by \citet{Lanzaetal03}, therefore only
a brief introduction is given here. It makes use of three active regions,
containing both cool spots and warm faculae, to fit the rotational modulation of
the TSI, and a uniformly distributed
background component to fit the 
variation of its mean level along the 11-yr cycle. The ratio of the area
of the faculae to that of the sunspots in each active region is fixed and
is indicated by $Q$. The brightness of the unperturbed photosphere is modelled by a 
quadratic limb-darkening law:  
\begin{equation}
I_{\rm u} (\mu) = \mathcal{C} (a_{\rm p} + b_{\rm p} \mu + c_{\rm p} \mu^{2}), 
\label{unp_int}
\end{equation}
with $a_{\rm p} = 0.36$, $b_{\rm p}= 0.84$ and $c_{\rm p} = -0.20$ and $\mathcal{C} 
\equiv [(1/4)(a_{\rm p} + 2 b_{\rm p}/3 + c_{\rm p}/2)]^{-1} = 4.88$ \citep[cf.][]{Lanzaetal03}. 

The contrast between the sunspot and the unperturbed photosphere, as measured
by the ratio of their specific intensities, $c_{\rm s} \equiv I_{\rm s}/I_{\rm u} = 0.67$,
is assumed fixed and independent of the position on the solar disk. The facular contrast
is assumed to vary as: $I_{\rm f}/I_{\rm u} = 1 +c_{\rm f} - c_{\rm f} \mu$, 
where $c_{\rm f} = 0.115$ and $\mu \equiv \cos \psi$, with
$\psi$ being the angle between the normal to a given surface
element and the line of sight, i.e., its limb angle.

The rotation
period $P_{\rm rot}$ is assumed as a free parameter ranging from 23.0 to 33.5 days,
and its value is determined, together
with the other free parameters, by minimizing the $\chi^{2}$. The inclination
of the rotation axis $i$ is held fixed at its true value during the fitting
process. The number of free parameters is therefore eleven. 

The timescale of evolution of the active region pattern on the Sun is significantly
shorter than the rotation period. Therefore, it is not possible to model
an entire rotation by adopting a fixed pattern of active regions. The
longest time interval that can be modelled with three stable active regions is
 14 days \citep[cf.][]{Lanzaetal03}.
Hence, the TSI time series is subdivided into subsets of
14 days, each of which is fitted with the model described above. 
The lag between the beginning
of two consecutively fitted time intervals is fixed 
at 7.0 days like in \citet{Lanzaetal03}, giving a total of 484 intervals when the gaps in the time
series are taken into account. 

The time series obtained by MOST and those to be obtained by COROT have an extension
ranging from $\sim 30$ to $\sim 150$ days. Therefore, they do not allow us to sample the 
whole range of  stellar irradiance variations along an activity cycle. For this reason, in the present work
we compute the best fits assuming as the reference irradiance value (i.e., that corresponding to
the unperturbed Sun) the maximum of the TSI along each of the 14-day subsets.
This approach differs from that of \citet{Lanzaetal03} where
the observed maximum of the TSI in the $1996.1-2001.6$ interval was adopted as a reference value. 
The present choice makes the role of the uniformly distributed background component
almost irrelevant and the total spotted area comes from the amplitude of the 
TSI rotational modulation along each 14-day interval. 

With this assumption for the reference value, 
it is found that the parameter $Q$ must be adjusted to the value $Q=9$, instead of the 
previously adopted $Q=10$  \citep[cf. ][]{Lanzaetal03}
to have residuals with a symmetric distribution around an average zero value. 
 The other parameters
adopted for the three-spot modelling are the same as \citet{Lanzaetal03}. 

The main limitation of the three-spot model is the low number of degrees
of freedom to map the surface of the Sun that cannot account for the very
complex active region configuration observed on the photosphere of our star,
especially during the phases of moderate or high activity.

\subsection{Maximum Entropy and Tikhonov models of the TSI variation}

To fit the variations produced by a complex pattern of active regions on the Sun, the best approach is
provided by regularized maps. The surface of the star is subdivided into
200 squared elements of side $18^{\circ}$ of latitude whose filling factors
are varied to reproduce the rotational modulation of the TSI. Specifically, we consider the 
filling factors of the unperturbed photosphere, the spotted photosphere
and the faculae that are indicated by $q_{\rm u}$, $q_{\rm s}$ and $q_{\rm f}$,
respectively, with $q_{\rm u} + q_{\rm s} + q_{\rm f} = 1$. The area ratio
between the faculae and the sunspots is fixed, i.e., we assume $q_{\rm f}/q_{\rm s} \equiv Q$.
In this way, the value of $q_{\rm s}$ for a given element fixes also the values of
$q_{\rm f}$ and $q_{\rm u}$ and can be used to specify the active region distribution.

By assuming the same dependence of the contrasts of spots and faculae adopted for the 
three-spot model (cf. Sect.~\ref{3spots}), 
the specific intensity of a surface element at limb angle $\psi$
and with a spot filling factor $q_{\rm s}$ can be written as:
\begin{equation}
I(\mu) = \left\{ 1 + \left[ c_{\rm s} - (1+Q) + Q (1 + c_{\rm f} - c_{\rm f} \mu )\right] q_{\rm s} \right\} I_{\rm u} (\mu),
\label{specintens}
\end{equation}
where the unperturbed specific intensity is given by Eq.~(\ref{unp_int}).

The contribution of the $i$-th element to the bolometric flux at time $t_{k}$ is: 
\begin{equation}
\Delta F_{ik} = A_{i} I[\mu_{i}(t_{k})] \mu_{i} (t_{k}) g[\mu_{i}(t_{k})],  
\label{fluxelem}
\end{equation}
where $A_{i}$ is the area of the element, $I$ is the specific intensity given by Eq.~(\ref{specintens}), 
$\mu_{i} (t_{k})\equiv \cos \psi_{i}(t_{k})$, with the limb angle $\psi_{i}$ of the 
element depending on its longitude, latitude and the time $t_{k}$, and $g(\mu)$ is the  visibility
function, i.e., $g(\mu) = 1$ for $\mu > 0$, $g(\mu )= 0$ for $\mu \leq 0$. To warrant a sufficient
precision in the computation of the disk-integrated flux, each surface element is further subdivided into 
smaller squared elements with side $1^{\circ}$ having the same filling factor $q_{s}$ of the initial element, thus
keeping the number of free parameters in our model equal to 200.
The bolometric flux at the time $t_{k}$ is given by: $F(t_{k}) = \sum_{i} \Delta F_{ik}$. 

The reduced $\chi^{2}$ statistic is defined as:
\begin{equation}
\chi^{2} = \frac{1}{M} \sum_{k=1}^{M} \frac{(F_{k}-D_{k})^{2}}{\sigma^{2}},
\label{chisq}
\end{equation}
where $F_{k}\equiv F(t_{k})$ is the computed flux and $D_{k}$ the observed flux at time $t_{k}$,
$\sigma$ the standard deviation of the TSI (equal to $2.0\times 10^{-5}$ in relative flux units),
and $M$ the number of data points in the fitted subset of the TSI series assumed to have a  length of 14 days
(cf. Sect.~\ref{3spots}).  

The regularized solutions are computed by a constrained minimization of an objective function that is 
a linear combination of the reduced $\chi^{2}$ and a regularizing function which
accounts for the a priori assumptions on the filling factor map, i.e.,
on the distribution of $q_{\rm s}$ over the surface of the star. Specifically,  the objective functions
are:
\begin{equation}
\mathcal{Q}_{\rm ME}= \chi^{2} -\lambda_{\rm ME} S, 
\label{QME}
\end{equation}
for the ME solution, and
\begin{equation}
\mathcal{Q}_{\rm T} = \chi^{2} + \lambda_{\rm T} T,
\label{QTik}
\end{equation}
for the T solution, where the reduced $\chi^{2}$ is given by Eq.~(\ref{chisq}), $\lambda_{\rm ME}$ and 
$\lambda_{\rm T}$ are Lagrange multipliers, and $S$ and $T$ are the ME and T regularizing functions
the expressions of which are given by \citet{Lanzaetal98}. The procedure of minimization, including the evaluation 
of the Lagrange
multipliers,  is described in detail by \citet{Lanzaetal98} to which we refer the reader.
Here we note only that the ME map is characterized by a minimum total spotted area because the 
value of $q_{\rm s}$ in each surface element is driven toward zero, whereas the T map is characterized
by a distribution of the filling factor as smooth as possible, i.e., the regularization
drives the map towards that of a uniformly spotted star. 

\section{Statistics to compare the models with the observations}

\subsection{Comparison of the modelled and observed TSI variations}

\citet{Krivovaetal03} and \citet{Wenzleretal05} compared the observed TSI variations with their models
by performing a linear regression between  measured and modelled values and found a correlation 
coefficient of 0.94. This approach can be extended to measure the performance of a given model to 
reproduce the variation of the TSI at different frequencies by means of  bivariate analysis 
\citep[cf., e.g.,][]{Priestley81}. We assume  a general linear relation between the 
observed  $D(t_{k})$ and the modelled TSI values $F(t_{k})$:
\begin{equation}
D(t_{k}) = \sum_{j=- \infty}^{\infty} f(t_{j}) F(t_{k-j}) + N(t_{k}),
\label{regress}  
\end{equation} 
where $f(t_{k})$ is the impulse response function  and $N(t_{k})$  the noise at time $t_{k}$, with the time instants
assumed to be uniformly spaced by $\Delta t$ (i.e., $t_{k} = k \Delta t$). Equation (\ref{regress}) means that the value of the observed TSI
at a given time is a linear combination of all the past, present and future values of the model, with the
function $f(t)$ giving a measure of the relative contributions at different time delays.
In this way,  relation (\ref{regress}) also takes into account a possible delayed (or anticipated)
response of our model.   

The relationship between the observed and the modelled time series is expressed
by the transfer function, that is the Fourier transform of the impulse response function: 
$\Gamma(\omega) \equiv \sum_{j=-\infty}^{\infty} f(j\Delta t) \exp(-i\omega j \Delta t)$, where $i = \sqrt{-1}$. An
ideal model would have $f(0) = 1$ and $f(t) = 0$, $\forall t \not= 0$, i.e., a gain $|\Gamma(\omega)| = 1$ and
a phase spectrum $\arg{\Gamma(\omega)}=0$ for all frequencies $\omega$. In other words, the model values   
would directly reproduce the observations without any delayed (or anticipated) contribution at all frequencies. 

The generalization of the linear correlation coefficient is 
the coherency which expresses the degree of linear correlation between the observed and modelled TSI as a
function of the frequency. When the 
modulus of the coherency is close to unity at all frequency, a linear best fit
reproduces well the correlation between the two quantities \citep[see][ for details]{Priestley81}.  

\subsection{Comparison of the total areas and longitude distributions of the active regions}

To assess the capability of the proposed modelling approaches, it is important to
compare the maps of the active region distributions obtained by light curve inversion with the
real distributions of the active regions on the Sun. In consideration of the limitations 
discussed in Sect.~\ref{model}, the comparison 
will concern the variation of the total active region area and the distribution of the active
regions in longitude. Since the ratio between the facular and the spot areas is fixed in our model,
we can consider the total spotted area and its longitudinal distribution as the model quantities to be 
compared with the observations. A daily record of the sunspot groups observed on the solar disk,
reporting their co-ordinates and areas, is provided by the extension of the  Greenwich Photoheliographic
Results (hereinafter GPR) made available by \citet{Hathaway06}. These data will be used for
our comparisons. 

It is  important to notice that
our models are based on subsets of the TSI time series that extend along half a solar rotation,  implying that points on the
solar surface having different longitudes are in view for  different fractions of the 14-day time interval. 
Therefore, when we compare the model active region areas with the observed GPR areas, we must take into account that the
model active regions contribute to the modulation of the TSI only for some fraction of the time. In other words, their
areas must be multiplied by this fraction to be consistently compared
with the observed areas during the same time interval. 

In view of the above considerations, we define the {\it visibility} of a point $P$ on the solar surface
along a given 14-day time interval as: 
\begin{equation}
v(P) = \frac{1}{(t_{2} - t_{1})} \int_{t_{1}}^{t_{2}} \mu(P, t) g[\mu(P, t)] dt,  
\end{equation}
where $t_{1}$ and $t_{2}$ are the initial and the final instants of the interval with $t_{2} - t_{1}= 14$ days,
and the functions $\mu$ and $g$ have already been defined above
\citep[see Eq. (2) of][ for the dependence of $\mu$ on the co-ordinates of the point $P$]{Lanzaetal03}. 

The comparison of the total area of the sunspots derived from our models with the observations
is straightforward once the element area values have been multiplied by their respective visibilities. 
The comparison of the longitudinal distributions is more complex. 
In the case of a discrete spot distribution,
we define the mean longitude of the active regions taking into account the circular nature
of the data, i.e., we associate to each active region of area $a_{j}$, longitude $\lambda_{j}$ and visibility $v_{j}$ a 
 vector $(a_{j} v_{j} \cos \lambda_{j}, a_{j} v_{j} \sin \lambda_{j})$ and then compute the resulting
vector normalizing it to the total area, i.e.:
\begin{equation}
{\bf R} \equiv \left( \frac{1}{\sum_{j} a_{j} v_{j}} \sum_{j} a_{j} v_{j} \cos \lambda_{j},
\frac{1}{\sum_{j} a_{j} v_{j}} \sum_{j} a_{j} v_{j} \sin \lambda_{j} \right),
\label{mean_long}
\end{equation}
where the summation is extended over the active regions. The mean longitude of the distribution is
given by the longitude toward which the resulting vector ${\bf R}$ points. In this way, we
can compare the mean longitude of the three model active regions with that of the  sunspot groups
observed during any given 14-day interval. In place of the visibility, we use the area averaged over the
14-day interval for the observed sunspot groups. 

The dispersion of the longitudes of the model active regions around their mean value 
is not a statistically meaningful quantity when only three active regions are 
adopted, hence we shall not consider it
in our discussion.

\begin{figure}[t]
\psfig{file=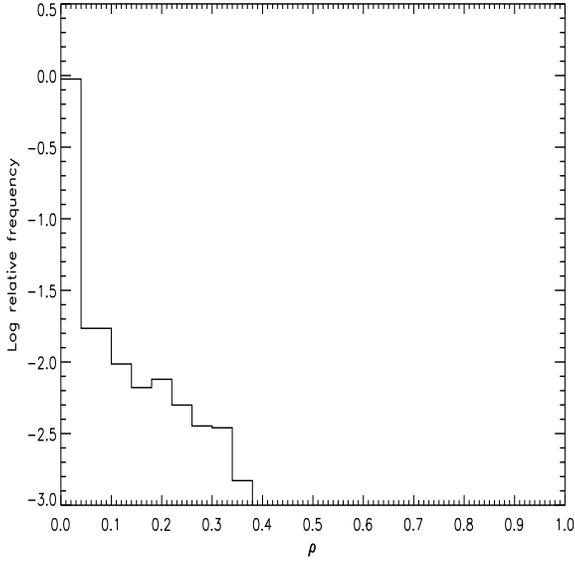,height=8cm,width=8cm} 
\caption{Histogram of the frequency distribution of the correlation coefficient $\rho$, as given by Eq.~(\ref{correl_coeff}),
between several observed longitudinal distributions of the sunspot group area on the Sun and the 
5040 random distributions obtained by the permutations of each of them, respectively.
}
\label{fig1}
\end{figure}

On the other hand, the longitudinal distributions of the spotted area obtained by the ME and the T maps are suitable for
a detailed comparison with the  longitudes of the sunspot groups observed during the same 14-day time
intervals. From the regularized maps, the spotted area in $18^{\circ}$ longitude bins can be easily
obtained and compared with the same quantity as derived from the observations. However, 
since the actual longitudinal resolution of our mapping method is certainly larger than
$18^{\circ}$, it is more appropriate to compare the distributions with a wider binning, say, 
$54^{\circ}$ or $72^{\circ}$. Taking into account the circular nature of our data,
a longitudinal distribution corresponds to a set of vectors, each having  
the direction of a given bin and  a modulus proportional to the total area in that bin. 
The correlation between two distributions $a$ and $b$ is measured by the correlation between the
two corresponding sets of vectors, e.g., $\{ {\bf V}^{a}_{j} \}$ and $\{ {\bf V}^{b}_{j} \}$, with $j = 1, ..., N$,
where $N$ is the number of bins. 
The correlation coefficient can be defined by specifying to the case of circular data (i.e., to two dimensions) 
the definition given in \citet{FisherLee86} and \citet{Fisheretal87}. 
More precisely, we define the quantity:
\begin{equation}
S({\bf V}^{a} {\bf V}^{b} )\equiv \det \left(
            \begin{array}{cc} 
              \sum_{j} X^{a}_{j} X^{b}_{j} & \sum_{j} Y^{a}_{j} X^{b}_{j} \\
              \sum_{j} X^{a}_{j} Y^{b}_{j} & \sum_{j} Y^{a}_{j} Y^{b}_{j}
            \end{array}   
\right),
\label{correlation}
\end{equation}
where $X^{a}_{j}$ and $Y^{a}_{j}$ are the components of vector ${\bf V}^{a}_{j}$ and $X^{b}_{j}$ and $Y^{b}_{j}$
those of vector ${\bf V}^{b}_{j}$. Then the correlation coefficient between the two sets of vectors, i.e.,
between the corresponding longitudinal distributions, is 
given by: 
\begin{equation}
\rho \equiv S({\bf V}^{a} {\bf V}^{b})/[ S({\bf V}^{a} {\bf V}^{a}) S({\bf V}^{b} {\bf V}^{b})]^{1/2},
\label{correl_coeff}
\end{equation}
with $0 \leq \rho \leq 1$ in the case of circular data. A correlation coefficient $\rho =1$ indicates
a perfect correspondence between the two distributions. For a given correlation coefficient $\hat{\rho} < 1$,
we can estimate the probability that $\rho$ is higher than $\hat{\rho}$ for the correlation between a given
distribution and a random distribution by means of a Monte Carlo method \citep{Fisheretal87}. For instance, 
in Fig.~\ref{fig1} we report the statistics of the values of $\rho$
obtained by correlating several observed distributions
with bins of $54^{\circ}$ with the 5040 random distributions obtained by the permutations of each initial one. The
probability of having $\rho \geq 0.4$ by a chance correlation is found to be lower than $\sim 0.1$\%.

\section{Results}
\label{results}

The best fits obtained with the three-spot model introduced in Sect.~\ref{3spots} 
and the corresponding $\chi^{2}$ values 
are comparable with those obtained by \citet{Lanzaetal03} and are not shown here.
The residuals have an approximately Gaussian distribution with a full width at half maximum
of $\sim 80$ ppm. 

The best fits with the ME and T regularizations are computed for the same  time intervals with 
the same choice for the unperturbed TSI value and $Q=9$ as for the three-spot model. However, the rotation period
is fixed at $P_{\rm rot} = 27.27$ days in order to avoid an excessively long computation time.
The Lagrange multipliers are set at $\lambda_{\rm ME} = 300$ and $\lambda_{\rm T} = 5 \times 10^{5}$
by the criterium that the distributions of the best fit residuals should not deviate from 
the almost Gaussian distributions obtained for $\lambda_{\rm ME}=0, \lambda_{\rm T}=0$ by more than
one standard deviation, i.e., $\sim 50$ ppm \citep[cf., e.g., ][]{Lanzaetal98}. 
We note that the present Lagrange multipliers are three orders of magnitude higher than those usually
adopted to model stellar light curves \citep[cf., e.g., ][]{Lanzaetal98,Lanzaetal02}. This is a consequence of the
much lower value of the standard deviation of the TSI measurements that makes the
trial values of the $\chi^{2}$ much higher than those in the stellar cases,
so that correspondingly higher Lagrange multipliers
are needed to make effective the regularizing terms in Eqs.~(\ref{QME}) and (\ref{QTik}).

\begin{figure*}
\unitlength1cm
\begin{minipage}[!t]{10cm}
\psfig{file=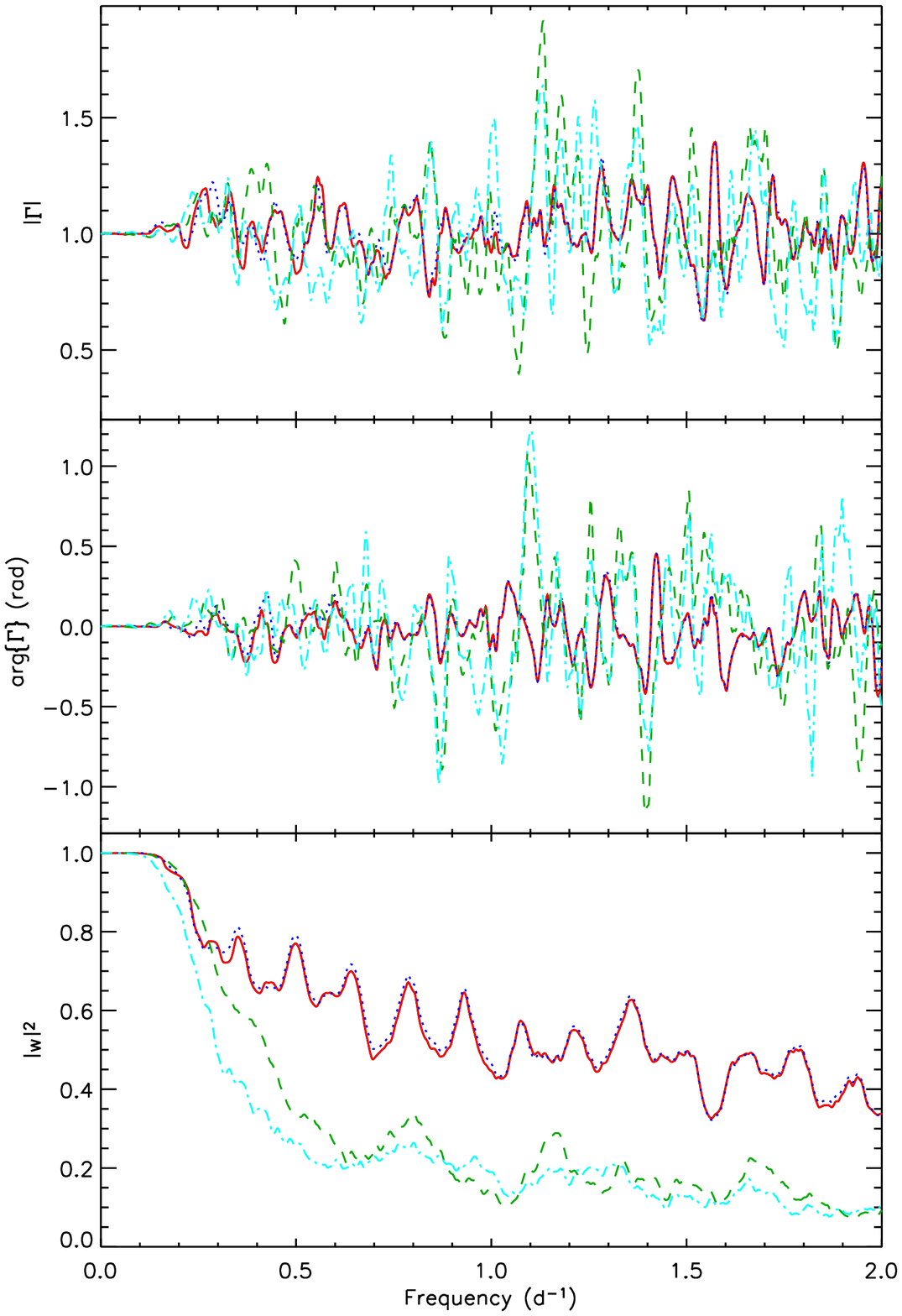,width=8.5cm}  
\end{minipage}
\hfill
\begin{minipage}[!t]{10cm}
\psfig{file=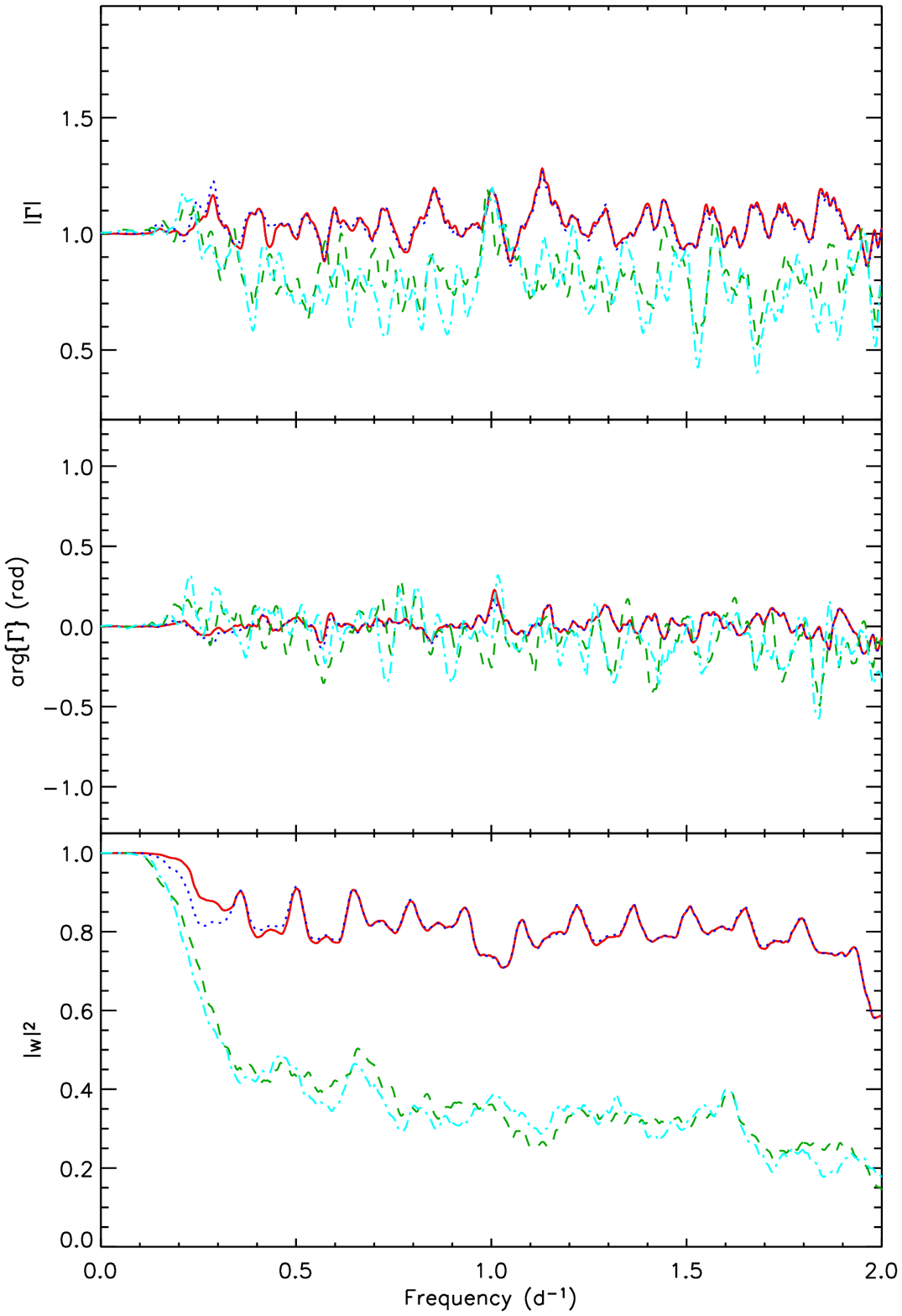,width=8.5cm}  
\end{minipage}
\caption{{\it From the upper to the bottom left panels:} The gain spectrum $|\Gamma|$, the phase spectrum $\arg{\{\Gamma\}}$ and the square of the modulus of the coherency $|w|^{2}$ versus frequency for our best fit time series 
versus the observed TSI time series during the interval ranging from 1996.25 to 1997.77.
Different linestyles and colors mark different modelling approaches: solid red -- ME; dotted  blue -- T; dashed green -- three spots; dash-dotted light blue -- two spots. {\it From the upper to the bottom right panels:} The same as the left panels for the
interval ranging from  1999.52 to 2001.00.}
\label{bivarminmax}
\end{figure*}

The best fits obtained with the ME and T regularizations are similar to those obtained with the
three-spot model and are not shown here. The $\chi^{2}$ values are usually comparable  
and are always lower than $12$, whereas in the case of the three-spot models $\chi^{2} > 12$
 in a few cases \citep[cf., e.g.,  Fig.~5 in ][]{Lanzaetal03}.
 In order to judge the performance of the
different approaches, we plot in Fig.~\ref{bivarminmax} the gain, the phase spectrum and the coherency
vs. frequency for the different model best fits with respect to the observed TSI for two time intervals, close
to the activity minimum and the maximum of cycle 23, respectively. In addition to the three-spot
models, we consider also models with only two spots. The average $\chi^{2}$ value of the three-spot models
is $\sim 30$\% lower than that of the two-spot models. We see that the gain and the coherency of the two-spot models
begin to depart from the unity already for frequencies of $\sim 0.1$ d$^{-1}$, while the three-spot
models are good up to $\sim 0.2$ d$^{-1}$ during the minimum
phases. Therefore, we shall not consider  the two-spot models any further.

 The ME and T models are closely comparable and perform significantly better than 
the discrete spot models, especially during the 
phases of higher activity, as indicated by the gain and the coherency close to unity and the phase angle close to zero
for frequencies lower than $\sim 0.25$ d$^{-1}$. For higher frequencies, the TSI variations
cannot be reproduced by any model based on the rotational modulation of a fixed pattern
of active regions because short-term irradiance fluctuations are ruled by the growth and decay of 
small active regions with a lifetime of a few days \citep[cf. ][]{Lanzaetal03}.
Small systematic differences in high-frequency residuals between consecutive best fits
separated by 7 days may be the cause of the oscillations observed in the plots of the coherencies
of the ME and T models for frequencies higher than $0.3$ d$^{-1}$. 

\begin{figure}[t]
\psfig{file=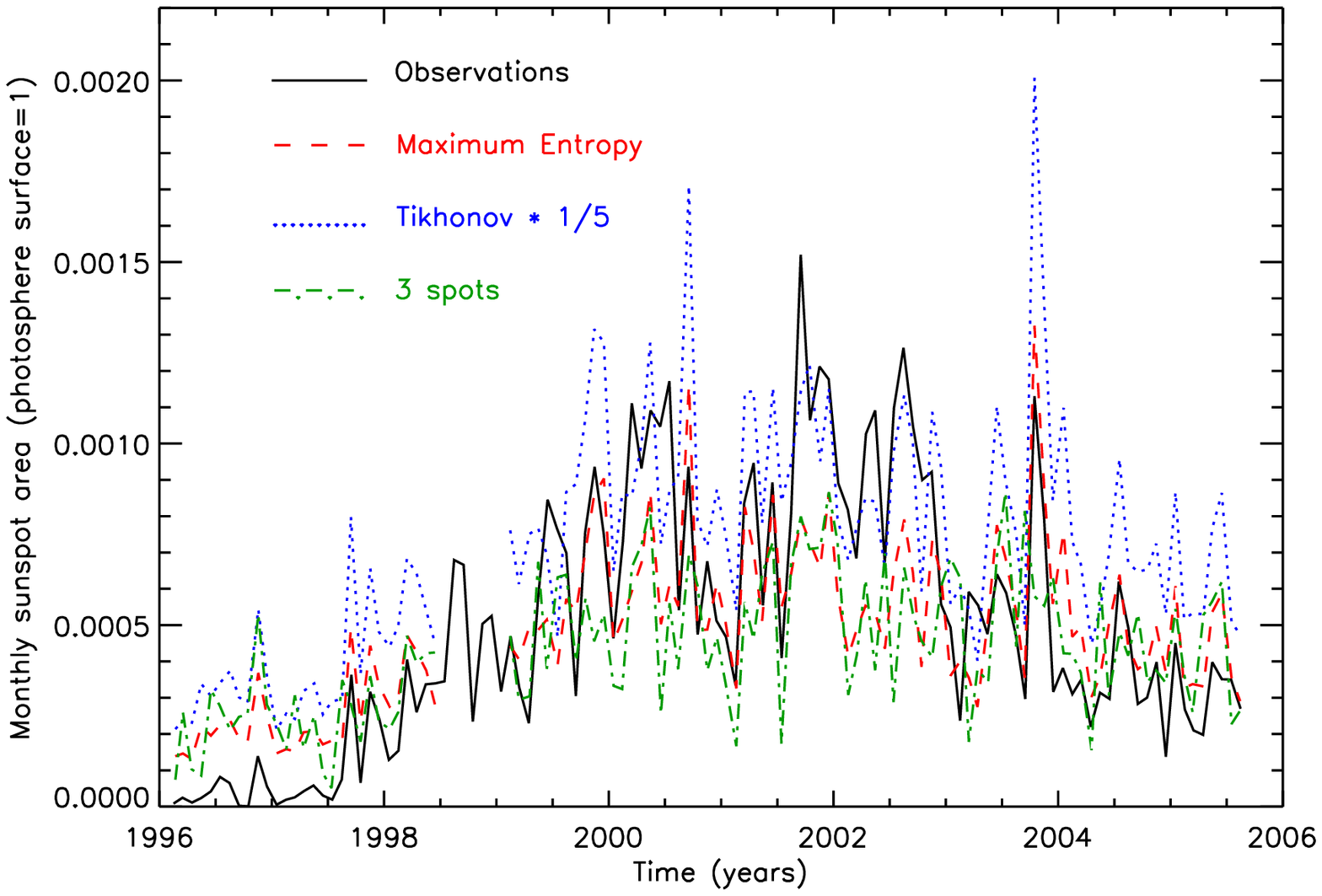,height=8cm,width=8cm} 
\caption{The monthly sunspot group area from the GPR and the averaged model spotted areas, in unit of the 
entire solar photosphere and corrected for the
effects of the visibility, vs. time along activity cycle 23. Note that the values of the areas obtained from the 
T models have been multiplied by $1/5$ to be plotted on the same scale. Different linestyles and colors  mark 
the various quantities: solid black -- observed GPR area; dashed red -- ME models; dotted blue -- T models; 
dash-dotted green -- three-spot models. 
}
\label{area}
\end{figure}

The total spotted areas obtained by the three-spot, the ME and the T models are compared in Fig.~\ref{area}
with the monthly averaged spotted area in the GPR. It is better to adopt for the comparison 
the mean model areas along an entire rotation. Thus, the individual values from four consecutive best fits,
separated by 7 days each from the other, are averaged.
Note that the T models give systematically higher values than the ME models because of their
regularizing assumption that smooths the area variation between neighbour surface elements
thus increasing the spotted area at high latitudes where its photometric effect is minimum. On the other
hand, the ME regularization tends to decrease  the spotted area in each surface
element as much as possible \citep[see ][ for details]{Lanzaetal98}.

The present results confirm and extend those presented by \citet{Lanzaetal03}, although it is
important to mention that the area values reported in that paper are different from ours because of the
different assumptions on the TSI reference value. The present model spotted
areas are systematically larger than the sunspot group area close to the minimum of activity when
the TSI modulation is dominated by the faculae. This happens because our models assume a fixed proportion between the
facular and the spotted areas. Hence, when they reproduce the rotational modulation at the cycle minimum they necessarily
introduce a certain amount of spotted area. The  modulation of the observed spotted area on time scales up
to $1-2$ yrs is generally well reproduced by our models, but the amplitude of the overall variation along 
the activity cycle 23 is somewhat underestimated. To interpret this result, it is useful to look at the 
plot of the TSI time series along cycle 23\footnote{{\small See Fig.~2.1 in
www.pmodwrc.ch/pmod.php?topic$=$tsi\\
/virgo/proj\_space\_virgo\#VIRGO\_Radiometry}} that 
shows that the amplitude of the TSI monthly variations has remained at approximately the same level from the end of 1999
up to the end of 2004. Since our model spotted areas are derived from the amplitude of the
rotational modulation of the TSI, 
the approximate constancy of their average values in the $2002-2004$ interval 
 reflects the frequent large dips in the TSI time series observed up to the end of 2004. As a matter of fact,
cycle 23 was a peculiar one, characterized by two maxima of comparable heights in 2000 and 2002-2003
which delayed the  descending phase of the cycle. Our models
do not reproduce the amplitude of such maxima, although they follow the shorter-term oscillations
of the spotted area occurred during each of them. 

\begin{figure}[t]
\psfig{file=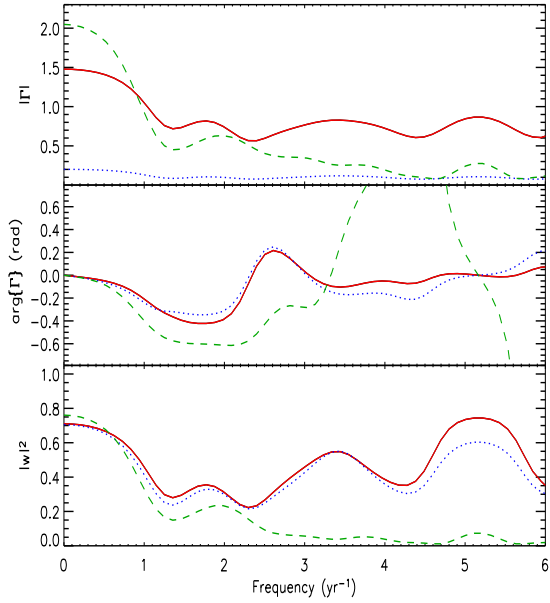,height=8cm,width=8cm} 
\caption{The gain spectrum $|\Gamma|$, the phase spectrum $\arg{\{\Gamma\}}$ 
and the square of the modulus of the coherency $|w|^{2}$ versus frequency for the
modelled area time series vs. the time series of the observed sunspot group area.
Different linestyles and colors refer to different modelling approaches: solid red -- ME; dotted
blue -- T; dashed green -- three spots. 
}
\label{area_bivar}
\end{figure}

These results are confirmed quantitatively by the bivariate analysis of the modelled vs. observed
sunspot group areas reported in Fig.~\ref{area_bivar}. The gain spectra show that the ratios of the 
modelled areas to the observed area depend on the frequency and they are different from unity 
even at the lowest frequencies. This depends on the different a priori assumptions made on the 
spot distributions adopted to fit the light curves, as discussed above in the case of 
the ME and T regularizations. 

The regularized models perform significantly better than the three-spot models with the ME
models showing the gain closer to unity.
 It is interesting to note that
the ME and T coherencies are lower in the frequency interval corresponding to time scales
between 0.25 and 0.8 yr, while their phase angles are significantly different from zero only for timescales
between 0.3 and 1.0 yr. For timescales shorter than about 0.25 years, both the coherencies and the phase angles
are similar to the values found for timescales longer than one year, indicating that the ME and T models
are reproducing the area variation quite well.  

\begin{figure}[t]
\psfig{file=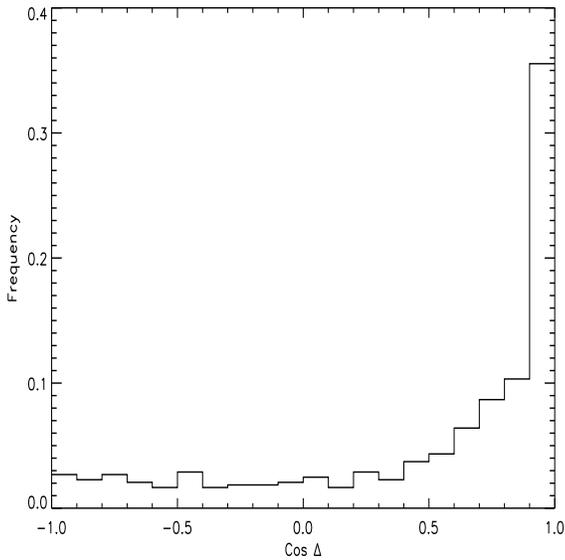,height=8cm,width=8cm} 
\caption{The frequency distribution of $\cos \Delta$ for the three-spot models, where $\Delta$ 
is the angle between the mean longitude of the three model active regions and the mean longitude
of the observed sunspot groups as given by Eq.~(\ref{mean_long}). The correction for the visibility
effects has been taken into account to compute the mean model longitudes. 
}
\label{long_corr_3spots}
\end{figure}

\begin{figure*}
\unitlength1cm
\begin{minipage}[t]{10cm}
\psfig{file=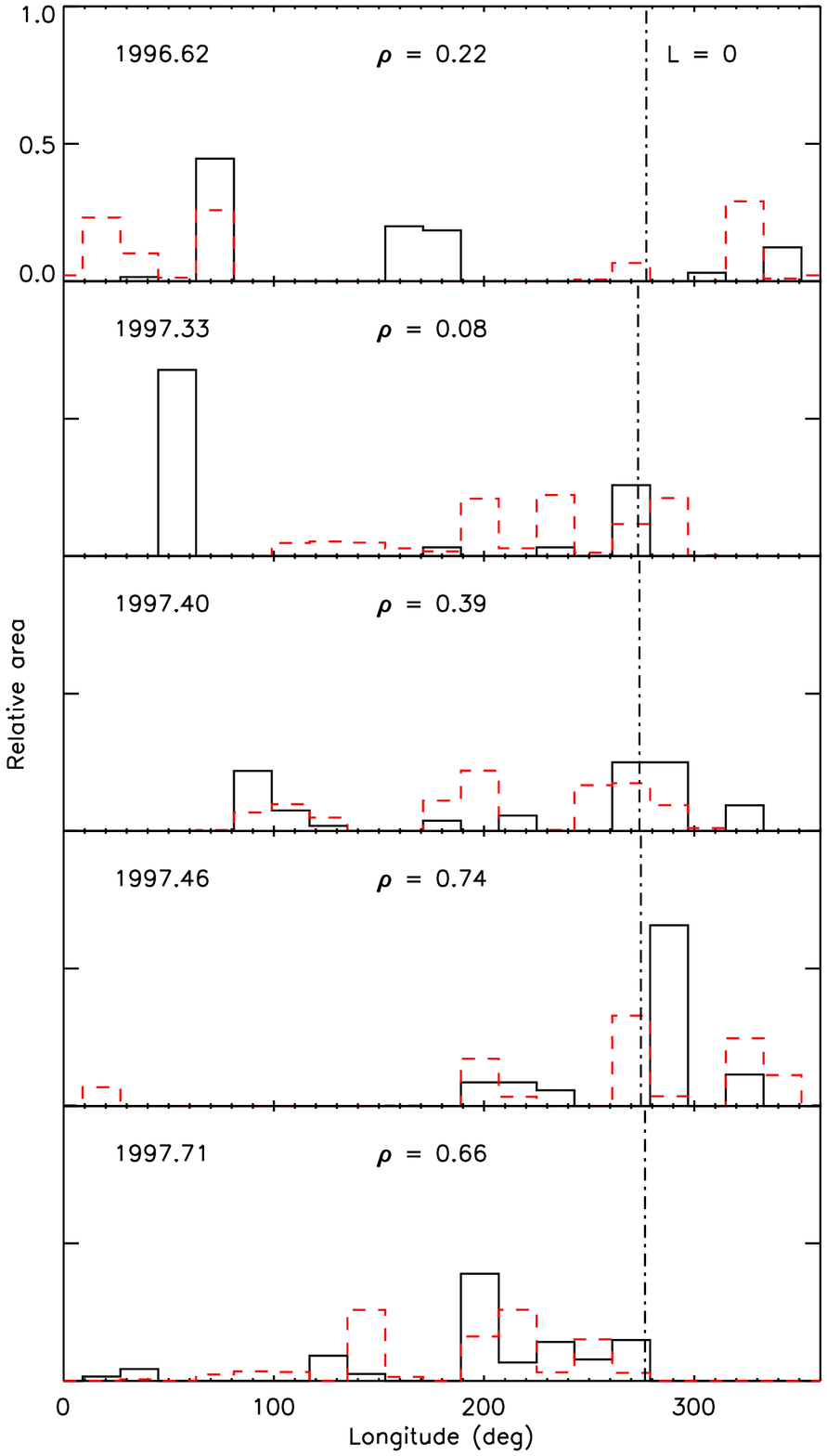,width=8.5cm}  
\end{minipage}
\hfill
\begin{minipage}[t]{10cm}
\psfig{file=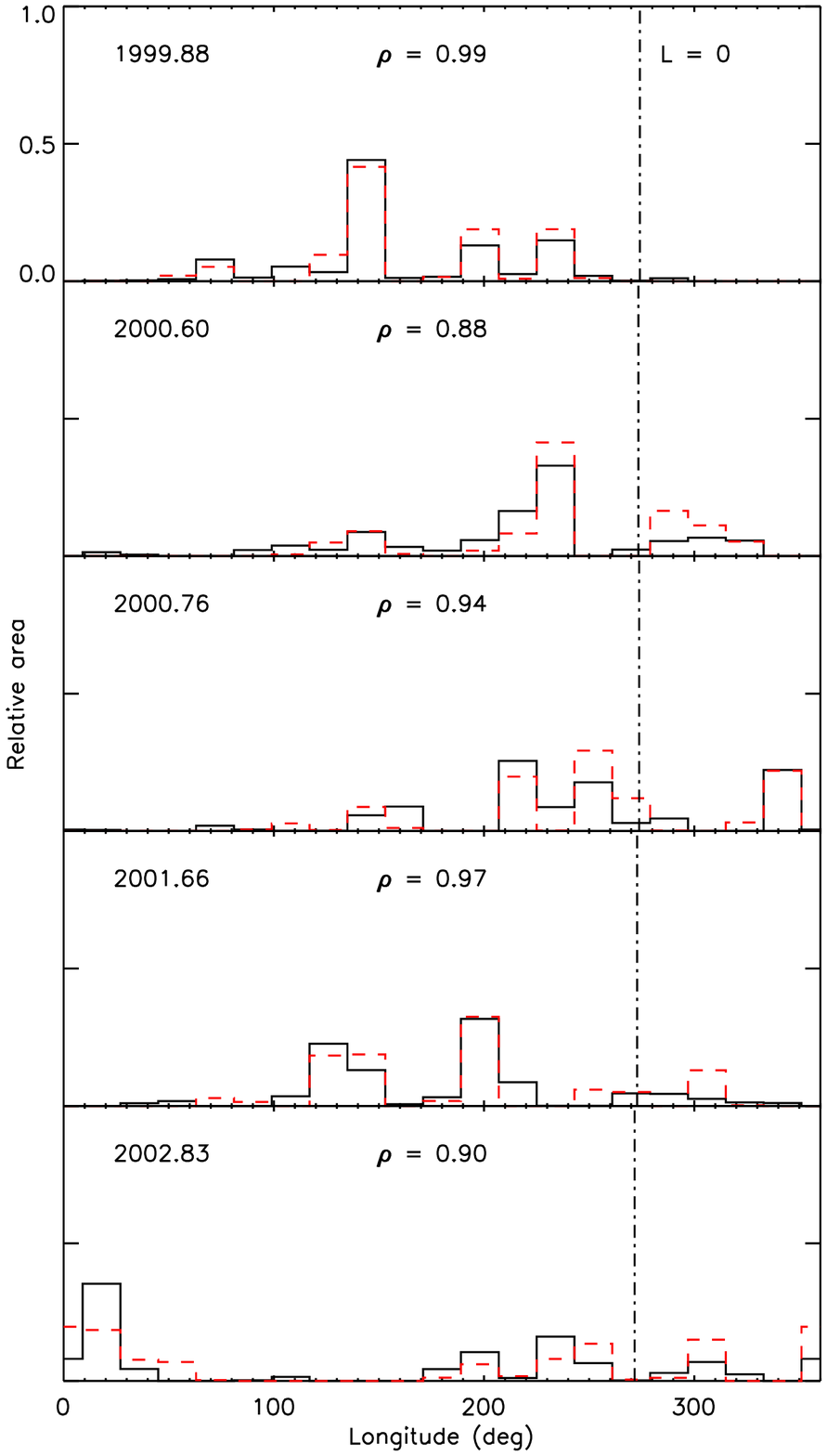,width=8.5cm}  
\end{minipage}
\caption{
The observed distributions of the sunspot group area (solid black line) and the corresponding ME
spot models (dashed red line) vs. longitude binned in $18^{\circ}$
intervals at the labelled epochs, grouping around the minimum of activity cycle
23 in the left column panels and around the maximum in the right column panels, respectively.
The modelled distributions are corrected for the effect of the visibility. Both the
observed and the modelled distributions are normalized to their total area, respectively.
The reference frame is the internal reference frame of our code (see the text).
 For comparison, the initial meridian of the Carrington reference
frame is marked in each panel by a vertically dot-dashed line, labelled $L=0$ in the uppermost panels.
The correlation coefficient
$\rho$ between the modelled and the  observed distributions, as 
defined by Eq.~(\ref{correl_coeff}) for longitude bins of $54^{\circ}$, is reported in each panel. 
}
\label{long_distr}
\end{figure*}

A discrete spot model based on three spots allows us to define the average longitude of
the spotted area distribution (cf. Eq.~(\ref{mean_long})) which can be compared with the
mean longitude of the distribution of the observed sunspot groups during the same 14-day time interval.
The comparison statistics are shown in Fig.~\ref{long_corr_3spots} and indicate that the model mean longitude
is within $60^{\circ}$ from the observed mean longitude in 68\% of the cases.
However, a discrete spot model based on three spots provides only a very rough description
of the longitudinal distribution of the spotted area. 
The ME and T models, assuming a continuous
distribution of the spotted area, give more information. We compare 
the distribution of the ME model spotted area  with the observed
distribution of the sunspot group area vs. longitude in the left and right panels of Fig.~\ref{long_distr} 
for selected epochs during the minimum and the maximum of activity cycle 23, 
respectively.  The origin of the longitude reference frame corresponds to the
meridian crossing the centre of the solar disk at 23$^{\rm h}$ 29$^{\rm m}$ 20$^{\rm s}$ UT
of 6 February 1996 for a geocentric observer. The adopted synodic rotation period of 
27.27 days makes the origin of our reference frame drift slowly with respect to 
the Carrington reference frame with, superposed, a small periodic oscillation
of amplitude $\sim 4^{\circ}$ and period of about one year.
The value of the area in each longitude bin in Fig.~\ref{long_distr} has been normalized to
the total spotted area of each distribution, i.e., the sum of the spot area over the bins
is always equal to the unity. Such a normalization has been adopted to avoid systematic effects
arising from the different total spotted areas of the model and the observations, as discussed above.
The correlation coefficients reported in Fig.~\ref{long_distr} are computed adopting a bin width of
$54^{\circ}$ with the area  in each bin  obtained by summing up 
the values of three consecutive bins of $18^{\circ}$. The optimal bin width has been chosen
by comparing the correlation coefficients for  bin widths of $36^{\circ}$, $54^{\circ}$ and 
$72^{\circ}$  and selecting the binning that gives  the higher resolution and the higher
correlation coefficient, on the average. Therefore, a bin width of $54^{\circ}$ corresponds
to the average longitudinal resolution of our ME spot modelling technique. Note, however, that
the original model bin width of $18^{\circ}$ has been retained for plotting the distributions
to show  the correlation between models and observations in more detail.
 
The distributions for the T models are usually broader than those of the ME models 
implying a lower degree of 
correlation betweeen the models and the observations and are not reproduced here. During the minimum of 
activity, the rotational modulation is dominated by the faculae, which implies that our models,
having a fixed proportion between the facular and the spotted areas in each surface element,
are often not capable of
retrieving the correct spot longitudinal distribution. Two cases of such poor agreement are
shown in the uppermost left panels of Fig.~\ref{long_distr} having a correlation coefficient 
$\rho$ lower than 0.25.
 The maximum of activity
is characterized by correlation coefficients usually higher than 0.8 indicating a
remarkably good agreement between the ME modelled and the observed sunspot group
area. 

\begin{figure}[t]
\psfig{file=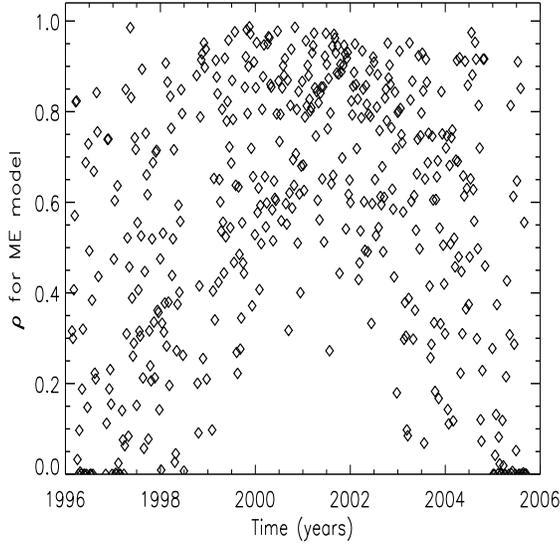,height=8cm,width=8cm,angle=0} 
\caption{The correlation coefficient $\rho$ between the ME spot model
and the observed longitudinal sunspot group distributions, as given by Eq.~(\ref{correl_coeff})
for longitude bins of $54^{\circ}$, vs. time. The model distributions are corrected for the effects
of the visibility. 
}
\label{rho_distr}
\end{figure}

The distribution of the correlation coefficient $\rho$ vs. time is plotted in
Fig.~\ref{rho_distr}.  The correlation is lower during the
phases of lower activity and it is remarkably good during the phases of 
intermediate or high activity with only a very few values lower than 0.4
during the maximum of cycle 23. As a matter of fact, we see from 
Figs.~\ref{fig1} and \ref{long_distr} that even a value as low as $\rho=0.4$ indicates a
significant degree of correlation between the modelled and the observed 
distributions. 

We have explored the effects of the variations of the model
parameters $Q$, $P_{\rm rot}$ and $i$,
by changing one parameter at a time while the others are kept fixed. 
We shall focus our discussion on
the ME models because they are those showing the best performance. 

It is interesting to consider models with
only dark spots, i.e., $Q=0$ because they are analogous to those usually adopted in the case of 
highly active solar-like stars \citep[cf., e.g., ][]{Messinaetal99,Lanzaetal02,Lanzaetal06}. 
The $\chi^{2}$ values of the best fits are systematically higher by a factor
of about $1.5-2.0$ with more than 10\% of the best fits having $\chi^{2}>15$. 
The total sunspot areas are somewhat smaller than in the reference  case ($Q=9$) because
smaller spots at the disk center can fit the dips in the TSI modulation, given that the
positive irradiance contribution from the faculae closer to the limb is missing. 
The longitudinal spot distributions are not reproduced as well as in the
reference case because the model spot longitudes are modified  to reproduce
the irradiance modulation due to the faculae which is systematically shifted in phase
with respect to that of the sunspots belonging to the same active region.

When $Q$ is increased beyond the value of the reference case, 
the $\chi^{2}$ values are comparable with those of the reference case up to
$Q \sim 15$, but the quality of the fits is significantly degraded for 
higher $Q$ values.
The total model spot areas increase with $Q$ because a larger sunspot
area close to the disk center is required to fit the TSI dips given the 
enhanced contribution of the faculae closer to the limb. 
The longitude distributions of the spotted area show $\rho$ values that
are comparable with those of the reference case up to $Q \sim 15$. For higher
values the longitudinal spot distribution is not satisfactorily reproduced by the 
models.  

The rotation period of the Sun cannot be accurately obtained by the analysis
of the rotational modulation of the TSI during the phases of moderately high
activity or close to the maximum of the 11-yr cycle \citep[cf., e.g., ][]{Lanzaetal03,Lanzaetal04}.
Therefore, we explored the effect of adopting a rotation period shorter by 20\%
or longer by 20\% than the optimal value of 27.27 days, respectively. In both cases, the
$\chi^{2}$ values of the best fits are not significantly affected and the 
variation of the total spotted area vs. time is well reproduced. However, the 
correlation coefficients $\rho$ between the modelled and the observed longitudinal distributions
of the spot area decrease significantly, in other words, the longitudinal resolution of
the model is degraded. The systematic error in longitude produced 
by a relative error $\Delta P/P_{\rm rot}$ in the rotation period is 
$\sim 180^{\circ} \times \Delta P/P_{\rm rot}$, so that the resolution is degraded by 
a factor of $\sim 1.5$ with respect to the reference case, when  an average
value of $50^{\circ}-60^{\circ}$ was found. 

Decreasing the inclination angle to fixed values of $60^{\circ}$ and $30^{\circ}$, respectively,
produces no significant degradation in the quality of the best fits as indicated by the
$\chi^{2}$ values that are comparable with those of the reference case. The total model spot
area does not change significantly for $i = 60^{\circ}$ and the longitudinal distributions
of the model spot area are well comparable with the observed ones, with no significant
degradation with respect to the reference case. 

\section{Discussion}

We have performed a detailed study of the capabilities of different spot
modelling techniques to retrieve information on the area variation
and longitudinal distribution of active regions through the modelling of
the rotational modulation of the solar irradiance along activity cycle 23.
Previous studies were limited to the Sun at minimum activity, i.e., characterized
by a geometrically simple configuration of the active regions \citep[e.g., ][]{Olahetal99},
or did not use quantitative statistical methods to study the correlation
between modelled and observed active region distributions 
\citep[e.g., ][]{Lanzaetal03}. Moreover, e.g., \citet{Olahetal99} and \citet{Catalanoetal98}
 did not model the optical flux modulation, but focussed on the flux coming from
the corona and the chromosphere, respectively. Those studies are, therefore, more
relevant for the modelling of the Sun as a star in the X-ray and UV spectral domains
\citep[e.g., ][]{Orlandoetal04}, rather than in  the optical band.

We have found that the ME models provide the best representation of the 
solar active region pattern in the photosphere. This is probably related to the
fact that active regions are localized features occupying a very small fraction of the solar surface,
a characteristic that  
matches the a priori assumption of ME models, i.e., that of concentrating the
brightness inhomogeneities in a number of surface elements as small
as possible \citep[cf., e.g., ][ and references therein]{Lanzaetal98}.

It is important to note that our approach cannot give the absolute
area of the sunspot groups but only the {\it variation} of the area
with respect to a reference level. As a matter of fact, the present
study demonstrates that the irradiance reference level need not be fixed, but it can be chosen as the maximum
of the TSI along each 14-day data subset. In other words, the amplitude of the 
rotational modulation along each data subset allows us to extract 
a good measure of the total spotted area, on the average. The  
longitudinal distribution of the active regions is
also well reproduced by our approach with an average resolution of 
$\sim 50^{\circ}-60^{\circ}$. This implies that a uniform distribution of
active regions or one characterized by variations on smaller longitudinal scales
cannot be reproduced by our models.  In the case of distant stars, the signal-to-noise
ratio will also affect the longitudinal resolution as discussed by, e.g., \citet{Lanzaetal06} in the 
case of ground-based observations. 

Forthcoming space-borne observations
will be limited to $4-5$ years  in the case of Kepler or  150 days in
the case of COROT. However, short-term cyclic variations of the spotted area are 
expected to be found,
 at least in some stars, by analogy with the 
Rieger cycles observed in the Sun. Specifically, near the maxima of  solar cycles
19 and 21, a periodicity of about 150-160 days was observed in the total
sunspot area \citep{Oliveretal98, KrivovaSolanki02} while other short-term periodicities
were sometimes detected in the flare occurrence rates
\citep[see, e.g., ][ and references therein]{Lou00}. Our ME models are particularly
suited to reveal variations on time scales of the order of 3-4 months, so
they can be applied to reveal Rieger cycles on other stars. 

The maps of the longitudinal distributions of active regions can be used to
look for preferential longitudes of spot activity, produced by complexes of 
activity analogous to those observed in the Sun that persist for time scales of 5-6 months.
On time scales comparable with stellar activity cycles, such as those 
of Kepler observations, it can be possible to trace the long-term
evolution of the active longitudes, possibly revealing their migration and
 connection with differential rotation, as claimed in the case of the Sun
by \citet{Usoskinetal05}. 

Moreover, the observations of stars harbouring extrasolar planets can be 
analysed to seek for a possible connection between the longitude of the 
active regions and the orbital phase of the planets, as suggested in the case of 
chromospheric plages by, e.g., \citet{Shkolnik05}. 

The results presented in Sect.~\ref{results} show that some 
parameters  can significantly affect the models. The
ratio of the facular to the sunspot area $Q$ affects not only the
value of the total spotted area but also the longitudinal
distribution of the active regions. The contribution of the 
faculae to the optical variations is probably relevant in  stars
that are brighter in the optical band at their maximum of chromospheric activity, as 
suggested by \citet{Radicketal98}. An estimate of their $Q$
may possibly be obtained from a comparison between the variations in the
optical band and those of the chromospheric fluxes in the cores of 
the Ca~II~H~\&~K or Ca infrared triplet lines, provided that there is a close association
between photospheric faculae and chromospheric plages.
Simultaneous observations in different passbands, such as those to be
obtained by COROT, may be used, at least in principle,  to constrain the
temperature of the surface inhomogeneities, e.g., according to the method devised  by
\citet{Messinaetal06}.  However, detailed simulations are needed to assess 
the capability of the proposed method with COROT spectrophotometry. 
In an ideal case, both techniques may be applied to constrain the values of 
$Q$ and of the contrast coefficients $c_{\rm s}$ and $c_{\rm f}$
\citep[see ][ for actual results derived by modelling the spectral solar irradiances]{Lanzaetal04}. 
In any case, the values of $Q$ and of the constrast 
coefficients can be varied within a certain range giving comparable best fits, especially
during phases of intermediate or maximum activity along the 11-yr cycle, 
because some degeneracy exists
between  $Q$, $c_{\rm s}$ and $c_{\rm f}$, cf. Eq. (\ref{specintens}).
Our results show that the timescales of the total area variations can be
identified even with significantly different values of $Q$, $c_{\rm s}$ and $c_{\rm f}$,
although the amplitudes of the area variations
depend on the parameters' values.
The systematic errors in the longitudinal 
distribution can be estimated by computing  models with different values
of $Q$ and comparing the results. 

For stars that become fainter at the maximum of their chromospheric activity
\citep{Radicketal98},
faculae are probably less important so that models with $Q=0$ may be adequate.
This is the case of the most active stars, as indicated by the results of
\citet{Lanzaetal06}. 

The rotational modulation of the chromospheric fluxes is useful to derive
the rotation period in the case of those stars having photospheric active regions
with a lifetime shorter than their rotation period, such as the Sun close to the 
maximum of activity \citep{Lanzaetal03,Lanzaetal04}. In principle, asteroseismic methods
can be used to obtain information of the rotation period and the inclination of the
rotation axis along the line of sight. In the case of the asteroseismic targets to be
observed by COROT, this approach will be possible for stars rotating with a 
period not exceeding a couple of weeks, but limited information will be retrievable at the
solar rotation rate \citep{Ballotetal06}. 

The inclination of the rotation axis is difficult to obtain from photometric data
only 
\citep[see, e.g., ][ for a parameter study of the MOST photometry of $\epsilon$ Eri]{Croll06}. 
In our models, we find a degeneracy between the inclination of the rotation
axis and the latitude of the active regions. However, the important point is 
that the longitudinal distribution of the active regions and the relative
variation of the total spotted area are not greatly affected by the uncertainty
on the inclination. The detection of planetary transits can provide a
constraint on the inclination of the rotation axis if the plane of the planetary
orbit is more or less normal to the stellar angular momentum as it is observed
in our solar system. 

The best fits provided by our models have residuals with standard deviations
of $50-100$ ppm  thanks to the fact that the solar flux variation
is dominated by the rotational modulation of a few active regions. The first 
results obtained by MOST suggest that the same kind of variability is characteristic
of other solar-like stars, also more active than the Sun, such as k$^{1}$ Ceti and
$\epsilon$ Eri \citep{Rucinskietal04,Crolletal06}. If confirmed by future observations,
this implies that most of the activity-induced variations in the light curves of 
solar-like stars can be fitted by means of our models, allowing a better efficiency
in the detection of planetary transits or light reflected from close-by giant planets.
 \citet{Lanzaetal03} showed 
a simple application of the method, but a detailed and comprehensive study
is still lacking and will be the subject of a forthcoming work. Here we only notice that
the residuals provided by the ME best fits are not Gaussian because of the 
influence of the regularization. Therefore, the three-spot model is probably better
suited to pre-process the light curves before applying transit searching algorithms that
are optimized for Gaussian white noise \citep[cf., e.g., ][]{Pontetal06}.  

\section{Conclusions}

We have tested different spot modelling approaches, applied to fit the
rotational modulation of the total solar irradiance, versus the observations of
the variation of the total sunspot area and longitudinal distributions. 
The comparison extends along almost an entire activity cycle
and is based on quantitative statistical methods that provide us with a description of the
model performance versus the timescale of the fitted variations. 
The ME models give the best agreement with the observations allowing us to reproduce the total sunspot area variation
on time scales ranging from a few months to the 11-yr cycle, although some systematic 
errors arise from the changing proportion of faculae and sunspots in active regions
along the activity cycle. The same approach is also capable of reproducing the longitudinal
distribution of the active regions with an average resolution of $\sim 60^{\circ}$, 
except during the phases of minimum activity when the rotational modulation is dominated
by solar faculae. 

The application of such a technique to the forthcoming high-precision
photometric time series by the space missions MOST, COROT and Kepler will provide us with a
quite detailed picture of photospheric magnetic activity in solar-like stars, especially
for those objects that have a level of activity comparable with the Sun, the study of 
which is not possible from the ground or through mapping techniques based on high-resolution
spectroscopy. Together with the chromatic information made
available by COROT for bright stars, such results will allow us to compare the solar
irradiance variations with those of other stars and to study the solar dynamo in 
the framework of the solar-stellar connection at the level of activity characteristic
of the present Sun. The modelling of more active stars, characterized by a faster
rotation, will allow us to study the evolution of magnetic activity versus age 
in solar analogues. 

Models to fit the stellar rotational modulation may  be useful to
enhance the efficiency of planetary transit detection techniques and to study the
connection between the presence of close-by giant planets and magnetic activity in
solar-like stars. 

\begin{acknowledgements}
     AFL and ASB wish to dedicate this work to the memory of 
     Marcello Rodon\`o, Professor of Astronomy in the University of
     Catania and Director of INAF-Catania Astrophysical Observatory.
     The authors wish to thank the anonymous Referee for a careful reading
of the manuscript. \\
\indent
The availability of  
the VIRGO/SoHO data on total solar irradiance and spectral irradiances
from the VIRGO Team through PMOD/WRC, (Davos, Switzerland) and of unpublished 
data from the VIRGO experiment on board of the ESA/NASA Mission SoHO
are gratefully acknowledged. \\
\indent
Research on stellar physics at Catania Astrophysical Observatory of the 
{\it INAF } ({\it Istituto Nazionale di Astrofisica}), and at
the Dept. of Physics and  Astronomy of Catania University is funded by MIUR 
({\it Ministero dell'Universit\`a e Ricerca})
and  {\it Regione Sicilia}, whose financial support is gratefully acknowledged. The extensive 
use of computer facilities at the Catania node of the Italian Astronet
Network is also gratefully acknowledged. 
\end{acknowledgements}

\end{document}